**Estimating lengths-of-stay of hospitalized COVID-19 patients using a non-parametric model: a case study in Galicia (Spain)**

**Authors**: Ana López-Cheda[1], María-Amalia Jácome*[1], Ricardo Cao[2], Pablo M. De Salazar[3]

**Affiliations**

1. Universidade da Coruña, CITIC, MODES, A Coruña, Spain

2. Universidade da Coruña, CITIC, ITMATI, MODES, A Coruña, Spain

3. Center for Communicable Disease Dynamics, Department of Epidemiology, Harvard School of Public Health, Boston, US

*****Corresponding author**: María-Amalia Jácome. Faculty of Science, Rúa da Fraga 10, 15008 A Coruña (Spain)

**Corresponding author email**:  maria.amalia.jacome@udc.es.

**ORCIDs**: Ana López-Cheda (https://orcid.org/0000-0002-3618-3246)[1], María-Amalia Jácome (https://orcid.org/0000-0001-7000-9623)[1], Ricardo Cao (https://orcid.org/0000-0001-8304-687X)[2], Pablo M De Salazar (https://orcid.org/0000-0002-8096-2001)[3]

**Abstract:** Estimating the lengths of stay of hospitalized COVID-19 patients is key for predicting the hospital beds demand and planning mitigation strategies, as overwhelming the healthcare systems has critical consequences for disease mortality. However, accurately mapping the time-to-event of hospital outcomes, such as the length-of-stay in the ICU, requires understanding patient trajectories while adjusting for covariates and observation bias, such as incomplete data. Standard methods, like the Kaplan-Meier estimator, require prior assumptions that are untenable given current knowledge. Using real-time surveillance data from the first weeks of the COVID-19 epidemic in Galicia (Spain), we aimed to model the time-to-event and event probabilities of patients hospitalized, without parametric priors and adjusting for individual covariates. We applied a nonparametric Mixture Cure Model and compared its performance in estimating hospital ward/ICU lengths-of-stay to the performances of commonly used methods to estimate survival. We showed that the proposed model outperformed standard approaches, providing more accurate ICU and hospital ward length-of-stay estimates. Finally, we applied our model estimates to simulate COVID-19 hospital demand using a Monte Carlo algorithm. We provided evidence that adjusting for sex, generally overlooked in prediction models, together with age is key for accurately forecasting hospital ward and ICU occupancy, as well as discharge or death outcomes.

**Author Contributions:** All authors contributed to the study and model design and formulation. ALC, MAJ and RC implemented the models and performed statistical analysis. All authors interpreted results and contributed to writing of the manuscript.

**Conflict of interest**:  'None declared'.

**Funding**: ALC was sponsored by the BEATRIZ GALINDO JUNIOR Spanish from MICINN (Ministerio de Ciencia, Innovación y Universidades) with reference BGP18/00154.  ALC, MAJ and RC acknowledge partial support by the MINECO (Ministerio de Economía y Competitividad) Grant MTM2014-52876-R (EU ERDF support included) and the MICINN Grant MTM2017-82724-R (EU ERDF support included) and partial support of Xunta de Galicia (Centro Singular de Investigación de Galicia accreditation ED431G 2019/01 and Grupos de Referencia Competitiva ED431C-2020-14 and

ED431C2016-015) and the European Union (European Regional Development Fund - ERDF). PMD is a current recipient of the Grant of Excellence for postdoctoral studies by the Ramón Areces Foundation.

**Word count**: 3594 words excluding tables, figures and references

**Introduction**

As of January 2021, SARS-CoV-2 transmission continues to increase in most countries worldwide [1], and in those countries where control has been achieved, resurgences are expected [2] before effective vaccines are widely available. Within the main challenges of the pandemic, overwhelming the healthcare systems has critical consequences on disease mortality [3]. Thus, understanding and predicting inpatient lengths of stay (LoS) and critical-care demand remain some of the major components of outbreak monitoring for decision-making and contingency planning.

Predicting hospital demand entails estimating a patient's LoS and the probability of hospital outcomes such as requiring admission to the Intensive Care Unit (ICU). Estimation of these variables is challenging as it requires investigating the patients' trajectories, and it must account for complexities in the data. For example, the LoS of some inpatients may be censored because the study ends before the patient leaves the hospital facility. The LoS of COVID-19 patients has been studied using parametric models [4], semiparametric methods [5], and nonparametric estimators [3, 6].

Parametric and semiparametric approaches are often preferred due to their simplicity and ease of interpretation, but they require the LoS to conform to a predefined fixed model. Estimations based on non-validated assumptions can be significantly biased. Thus, nonparametric approaches, which do not require model assumptions, should be used when estimating COVID-19 LoS in the absence of solid knowledge.

The Kaplan-Meier (KM) estimator [8] is the simplest and most frequent nonparametric estimator in medical time-to-event data. It assumes that all patients with missing outcomes would experience the event in the end. This assumption applies when analyzing the duration of hospitalization, that is, the total time in the institution of the hospital (which includes time in hospital ward and time in ICU), as all patients leave the hospital eventually. However, this assumption does not apply to a patient's LoS in the hospital ward until some of the potential outcomes, like admission to the ICU or until death, as not all patients need admission to the ICU or die. Thus, the KM estimator should not be used to estimate those LoS, as it is wrongly specified. Alternatively, Mixture Cure Models (MCM) [9] account for the situations when a proportion of individuals will not experience the event being analyzed.

Here, we propose a nonparametric Mixture Cure Model (NP-MCM) for estimating the LoS until specific events that are not experienced by all the patients. Specifically, we computed the following 5 lengths-of-stay: LoS in hospital ward until admission to ICU, LoS in hospital ward until discharge from hospital ward, LoS in hospital ward until death in hospital ward, LoS in ICU until discharge from ICU; and LoS in ICU until death in ICU. Note that the KM estimator would not be biased to model the LoS until discharge if all status on discharge were gathered as a composite outcome. Although only the first LoS is necessary to model ICU demand, the other LoS are also of interest, as they are useful to estimate the conditional probability that a patient experiences each of those events according to the corresponding observed LoS. Last, we also estimated the probability of each event.

First, to illustrate how our model improves data fitting, we compared the NP-MCM to the KM estimator (which assumes that all the individuals will experience the event) and to the empirical (E) estimator (which discards all observations which event is not observed) for a dataset of COVID-19 patients from the first weeks of the epidemic in Spain. We further simulated inpatient and critical care incidence during an outbreak, along with the final outcome (discharge or death), using the estimated values, and adjusting for age and sex. Our model shows the importance of these individual variables for predicting hospital demand during transmission.

**Materials and Methods**

*Data source*

The dataset contains 10454 confirmed COVID-19 cases reported in Galicia (North-West Spain), from March 6$^{th}$ to May 7$^{th}$, 2020. Since not all of them required hospitalization, our study only included the 2453 patients who were admitted in hospital/ICU during that period. For the patients with several hospital ward (HW) and/or ICU admissions, the considered LoS was the first recorded one, that is, the number of days from their first entrance in HW/ICU until they experienced the outcome of interest. Data was provided by the regional public health authority, Dirección Xeral de Saúde Pública [10]. The data included information on age and sex; the dates of COVID-19 diagnosis, admission to the hospital and/or ICU; and the patient's last known clinical status. A summary of the dataset can be found in the Supplementary Material Section S1, see also [11] for other results on this dataset.

*Model formulation*

Mixture cure models [9], a special case of cure models [12], explicitly model survival as a mixture of two types of patients: those who will experience the final outcome and those who will not (known as "cured"). Note that here a "cured" individual is defined as being free of experiencing the event of interest, not necessarily cured in medical terms. The goal of MCM is to estimate the probability of experiencing the event and the distribution of the time to the event. The model is formulated as follows.

Let us denote Y as the time to the event of interest (admission to ICU, death, or discharge), with survival function S(t) = P(Y > t). Let p = P(Y < ∞) be the probability that the event will happen, and $S_0(t)$ = P(Y > t | Y < ∞) be the survival function of the individuals experiencing the event. MCM write the survival function as S(t) = (1 − p) + p $S_0$(t). Then the probability of the event, p, and the survival function of the time-to-event, $S_0$(t), can be estimated using a proper estimator of the survival function, S(t), and the relations:

$$p = 1 - S(\infty) \text{ and } S_0(t) = \frac{S(t) - (1-p)}{p} \quad \text{(eq1)}$$

When there is a group of patients known not to experience the event, the survival function S(t) can be estimated nonparametrically as follows [13]:

$$\hat{S}(t) = \prod_{t_i < t} \left(1 - \frac{d_i}{n - i + 1 + \sum_{j=1}^{i} x_j}\right), \quad \text{(eq 2)}$$

where $t_i$ is the observed time-to-event of all the patients, $d_i$ indicates if the event was observed, and $x_i$ the indicator of whether the event was not observed because it is known it will never happen. To note, this estimator reduces to the well-known KM estimator in a classical time-to-event analysis when the event happens for all patients or when there is no way to identify the patients who will not experience the event ever.

The estimator of S(t) in (eq2) is computed with R software [14] and used to estimate the probability of the event, p, and the time-to-event survival function $S_0$(t) using the relationships in (eq1) for the five LoS aforementioned in the Introduction. Details on each LoS, along with an R script for the computation of the different estimators, can be found in the Supplementary Material.

The NP-MCM survival estimator of $S_0(t)$ is compared to the KM estimator computed with two different datasets: (i) the complete set of observations, considering all the patients who are known not experience the event as simply right censored regardless if they might experience it in the future or not (complete KM), and (ii) a reduced dataset, dismissing the patients who will not ever experience the event (reduced KM). The empirical E estimator, which considers only patients whose final event is observed and disregards the right censored observations, has also been considered. The NP-MCM estimator of the probability, p, of the event was computed using the estimator of S(t) in (eq2) and the relationships in (eq1). The empirical E estimator of p, given by the ratio between the number of observed events and the total number of patients, was computed to motivate the proposed NP-MCM estimator of p (see Supplementary Material Section S2 for details).

The NP-MCM estimator of S(t) in (eq2), the E estimator and the KM estimator do not incorporate possible covariate effects, such as sex and age. When the final outcome is experienced by all the patients, the extension of the KM estimator to handle covariates is the generalized product-limit estimator [15] of the conditional survival function, S(t|x). When the final outcome is not experienced by all the patients ('cured' individuals, all unidentified) the incorporation of covariates in the estimation of the probability of the final outcome p(x) and the distribution of the times until the event $S_0(t|x)$ has been studied recently [16-18] and implemented in the R package npcure [19], which also performs significance tests for the probability of the event p(x). When the final outcome is not experienced by all the patients ('cured' individuals, some of them identified as it happens for our COVID-19 data), the extension of these methods for the NP-MCM model to estimate p(x) and $S_0(t|x)$ has been recently addressed [13], where evidence of the superiority of the NP-MCM over the traditional methods is shown. These conditional estimators of S(t|x) [15], p(x) and $S_0(t|x)$ [13,16-18] can handle continuous covariates such as age, using the information from all the individuals to provide estimates for one single value, e.g., 40 years. Ignoring the effect of age and sex on these estimates can produce important bias in the statistical analysis.

*COVID-19 outbreak simulation model*

We further simulated a COVID-19 outbreak based on the NP-MCM estimates of the 5 lengths-of-stay considered, with two different models: 1) the simplest possible where the distributions of the lengths of stay and probabilities of moving from one state (hospital ward, ICU) to another (hospital ward, ICU, death, discharge) do not depend on individual covariates; and 2) a more realistic one with the LoS and transition probabilities depending on the age and sex. For the ease of computation, the simulated LoS were not generated directly from the NP-MCM estimates but from the parametric distribution that best fitted the NP-MCM estimates, specifically, Weibull distributions (see Supplementary **Figure S4** for the NP-MCM estimates and their corresponding Weibull counterparts).

The simulated outbreak consisted of N = 1000 infected individuals. For the *i*-th infected individual i = 1, …, N we simulated the sex $G_i$ (0 = male, 1 = female) and the age $A_i$ (years) using the real distributions of the reported COVID-19 cases in Galicia on May 7[th], 2020 (Supplementary **Table S1**). As not all the infected individuals required hospitalization, let $H \subset \{1, ..., N\}$ be the infected subjects admitted to the hospital. The trajectory of every hospitalized patient $i \in H$ was obtained by simulating the transitions between states (hospital ward, ICU, discharge, death) using the NP-MCM estimated probabilities. The times in each state were simulated from the Weibull distributions that best fitted the NP-MCM estimates, both conditional and unconditional on the age and sex of the patient (see Supplementary Material Section S4 for further details; Supplementary **Figure S4**; **Figures S5** and **S2**, **Table S3**). From the evolution of all the hospitalized patients, it is straightforward to compute the number of patients in every state. We simulated 1000 outbreaks of N = 1000 infected people, so the mean number of patients in a hospital ward, in the ICU, dead and discharged can be approximated by a Monte Carlo simulation for each day as a function of time. For supporting the goodness of fit of the conditional model that considers the age and sex of the patient, the real number of inpatients in the COVID-19 dataset has been taken as reference, rescaled to N = 1000 infected people.

## Results

We first compared the estimates of the LoS using the NP-MCM estimator with the E estimator, and the KM estimator with the complete and reduced dataset.

When an event happens for all patients (a.s "leave the hospital", when all status on discharge gathered as a composite outcome), KM is not biased and coincides with the NP-MCM, both of them represented with the one single line in **Figure 1** (top left) and Supplementary **Figure S2** (top left).The NP-MCM and KM estimators consider the n = 2453 hospitalized patients of which 2142 experienced the event (they left the hospital within the study's timeframe). The E estimator considers only the 2142 patients who left the hospital, disregarding the information from the 311 patients still in hospital. This biases the E estimate towards shorter LoS, as hospitalized patients with longer LoS are not be included in the estimation.

Further, when the final outcome is experienced by only a proportion of patients ("admission to ICU", "death", "discharge"), the KM (with both the complete and reduced samples) overestimates the time-to-event showing longer LoS than the NP-MCM. The E estimator underestimates the time-to-event due to right censoring, showing shorter values of LoS as it only takes into account patients who experienced the event. The NP-MCM estimates do not suffer from a similar bias [13]. In fact, this is one of the advantages of using these methods. Interestingly, we found small differences between the NP-MCM estimates and the E estimates (**Figure 1** and **Supplementary Figure S2**). The reason might be that the distribution of the LoS of the dismissed patients in the E estimation (never admitted to ICU) is similar to that of the patients who required ICU.

Importantly, for the probability of the medical event (admission from HW to ICU, death, discharge from HW or ICU) we showed that not correcting for right censoring (i.e., using the E estimator with only individuals with the observed outcome) underestimates the true probability, as the event of the right-censored individuals could be recorded later in time. The NP-MCM can adjust to right censoring, providing more accurate estimates. This can be seen when comparing individual probabilities using NP-MCM and E estimators (**Table 1**). Note that in **Table 1**, the probabilities of mutually exclusive outcomes are not equal to 1. This is because the final outcome (death or discharge) of 42 inpatients still in ICU at the end of the study remains unknown, as detailed in Sections S2.4 and S2.5. This inconsistency of the E estimates is partially corrected by the NP-MCM estimate.

Then, we used the NP-MCM estimator to assess if age and sex could play a role in the estimates of the time of hospitalization (both hospital ward and ICU) and the time in ICU. **Figure 2** shows that the LoS differ significantly between male and female patients, and between middle-aged (40y) and older (70y) patients. Particularly, we found that middle-aged female patients showed shorter LoS in both the institution of hospital and the ICU, while older females showed longer LoS in the ICU (but not in the hospital) compared to their male counterparts.

Finally, we implemented a COVID-19 outbreak simulation of N = 1000 infected individuals, using the NP-MCM estimates for the COVID-19 patients in Galicia (Spain) and accounting for age and sex heterogeneity in the LoS. **Figure 3** shows the difference in the simulated number of inpatients between considering age and sex (conditional model) or not considering age and sex of the patient (unconditional model). The higher the curve the more inpatients the model predicts. The real number of inpatients in the COVID-19 dataset, rescaled to N = 1000 infected people, has been added as reference. We found no large differences in the expected number of patients dead or discharged, regardless of age and sex are considered or not. Likewise, no large differences were found in the number of patients in ICU during the first month (until day 30 approximately) of the epidemic. However, after one month, the unconditional model tends to underestimate the number of patients in ICU, whereas the conditional estimation is closer to the real number of ICU inpatients in the COVID-19 dataset. The findings for the estimated number of patients in the hospital ward are similar for a period of 2 months. As a consequence, if prediction of hospital ward and ICU beds demand is estimated disregarding the sex and age of the patients, those predictions will be clearly underestimated, mainly in the case of ICU capacity. The consequences of this wrong forecast of hospital ward and ICU occupancy is shown in **Figure 4**. For a range of possible capacities (15-90 beds in HW, 5-15 beds in ICU), **Figure 4** shows the number of days in which the predicted number of patients exceeds the capacity. As expected, there is a decreasing trend, since the lower capacity the more days with excess demand. If age and sex are disregarded for making predictions (unconditional model), **Figure 4 (right)** suggests that 11 ICU beds are enough to avoid overload in the ICU. However, that ICU capacity will be exceeded for 18 days, and the available ICU beds should be set to 12 instead to prevent overwhelming of the ICU. Similar conclusions can be drawn when predicting HW beds demand in **Figure 4 (left)**. These discrepancies, simulated for N = 1000 infected people, would worsen as the incidence increases. In summary, while no large differences are observed in the predicted number of deaths and discharges, the conditional model gives more accurate estimates of the HW and ICU beds demand. This leads to a reduction in the number of days when the number of inpatients exceeds the HW and ICU capacity.

## Discussion

We applied a NP-MCM to estimate the time-to-event and event probabilities, including length-of-stay in hospital ward and ICU and time to death or discharge. In this work, we demonstrate how the LoS of hospitalized COVID-19 patients evolve over time, given age and sex distributions matching those from our database. The proposed model outperformed the KM and the empirical estimators when the outcome is not experienced by all patients. Importantly, the model can be adjusted for the use of covariates, which is significant when conditioning for known heterogeneity in estimating LoS. Particularly, our analysis demonstrates that adjusting for age and sex is crucial in accurately understanding ICU LoS and, in turn, forecasting bed demand.

Often studies with incomplete follow-up data on patients choose to exclude these patients from the study altogether, which yields biased estimates [20]. Moreover, when forecasting hospital demand in (near) real time, information related to the most recent cases is not available, which again leads to right censored data. We showed that the empirical estimator introduced significant bias toward longer LoS from HW admission to ICU admission because it ignores patients in the HW without ICU admission. Alternately, the KM estimator yields biased estimates towards longer stays as well when the event is not experienced by all patients. The reason is that the KM estimator assumes that if the follow-up time was long enough, much longer stays would be observed. Therefore, by comparing the NP-MCM against the KM estimate, we show how biased the results are when using the KM estimate. This comparison would support the use of the nonparametric cure model approach, which diminishes the problem that occurs when not all subjects experience the event.

Our findings are consistent with previous work: a recent systematic review has shown that median overall hospital stays ranged from 4 to 21 days outside of China [21], while our model estimated a median overall hospital stay of 11 days (IQR 7 – 19); the LoS for patients who died in the HW was generally shorter than those discharged alive (median 7 days and 10 days respectively). In contrast, our estimates show a different trend with regards to ICU LoS, with similar median estimates for both death and discharged (15 days vs 14 days), again consistent with that reviewed by Rees et al [21]. Of note, to our knowledge only two studies have adjusted LoS by age, all showing increased LoS for increased age, which is consistent with our findings [22, 4]. Furthermore, as far as we know this is the first study showing the influence of sex in the LoS, which has important implications for predicting hospital demand (**Figure 2**). With regards to prediction models, some approach adjust estimates based on age [23, 24], while sex has generally been overlooked in hospital demand forecasting [25, 23, 7].

Noteworthy, multi-state models [26-29] could seem an alternative method. Yet, multi-state cure models, in which transitions into one or more of the states cannot occur for a fraction of the population, are quite recent and the scarce literature is limited to semi and parametric models [30, 31]. Applying a nonparametric multi-state cure model is not straightforward, since there is no available literature related to this model. As a consequence, multi-state cure models were not used in this paper, but remain as a potential alternative approach.

Last, we would like to highlight key limitations of our model: the lack of a parametric function limits interpretability to a great extent and complicates handling several covariates simultaneously [32]. Regarding the application of MCM, there must be good evidence that some individuals in the population will never experience the event of interest and the follow-up time must be long enough [33]. To date, there has not been developed a reliable method for computing uncertainty (confidence bands) using MCM estimators, which remains a limitation of our approach Finally, data on patient comorbidities, which likely represents an important source of heterogeneity in the LoS, were not available for the analysis. Thus, more accurate estimates of the different LoS can be obtained if more complete datasets are available.

In summary, we implemented a NP-MCM that improved the standard survival methodology when estimating LoS until final outcomes that will not happen for all patients. We also found that the LoS in the ICU is sensitive to age and sex, which in turn is relevant when forecasting hospital demand in real-time for public health response. We believe our proposed approach can be easily implemented in other settings and can provide more accurate estimates of COVID-19 health demand compared to previous methods.

**Acknowledgments:** We thank the Dirección Xeral de Saúde Pública, Xunta de Galicia for providing the database.

**Conflict of Interest**: None

**Data Availability Statement** The data that support the findings of the study is available as aggregates in Table S1 of the supplementary material. Line list data is available upon request to the General Directorate of Public Health of the Autonomous Government of Galicia in Spain (Dirección Xeral de Saúde Pública, Xunta de Galicia).

**Funding**: ALC was sponsored by the BEATRIZ GALINDO JUNIOR Spanish from MICINN (Ministerio de Ciencia, Innovación y Universidades) with reference BGP18/00154. ALC, MAJ and RC acknowledge partial support by the MINECO (Ministerio de Economía y Competitividad) Grant MTM2014-52876-R (EU ERDF support included) and the MICINN Grant MTM2017-82724-R (EU ERDF support included) and partial support of Xunta de Galicia (Centro Singular de Investigación de Galicia accreditation ED431G 2019/01 and Grupos de Referencia Competitiva ED431C-2020-14 and ED431C2016-015) and the European Union (European Regional Development Fund - ERDF). PMD is a current recipient of the Grant of Excellence for postdoctoral studies by the Ramón Areces Foundation.

# References

1. **World Health Organization (WHO)**. Coronavirus disease (COVID-19) weekly Epidemiological Update 109 January 2021. https://www.who.int/publications/m/item/weekly-epidemiological-update---12-january-2021

2. **Kissler SM, *et al*.** Projecting the transmission dynamics of SARS-CoV2-through the postpandemic period. *Science.* 2020; **368**: 860-868. DOI: 10.1126/science.abb5793

3. **Grasselli G, Pesenti A, Cecconi M**. Critical care utilization for the COVID-19 outbreak in Lombardy, Italy: Early experience and forecast during an emergency response. *Journal of the American Medical Association.* 2020; **323**: 1545–1546. DOI:10.1001/jama.2020.4031

4. **Lewnard JA, *et al***. Incidence, clinical outcomes, and transmission dynamics of severe coronavirus disease 2019 in California and Washington: prospective cohort study. *British Medical Journal.* 2020; **369**: m2205. DOI: 10.1136/bmj.m1923

5. **Thai P, *et al***. Factors associated with the duration of hospitalization among COVID-19 patients in Vietnam: A survival analysis. *Epidemiology & Infection.* 2020; **148:** E114. DOI: 10.1017/S0950268820001259

6. **Wang Z, *et al***. Survival analysis of hospital length of stay of novel coronavirus (COVID-19) pneumonia patients in Sichuan, China. *medRxiv.* 2020; 040720057299. DOI: 10.1101/2020.04.07.20057299

7. **Grasselli G, *et al***. Risk factors associated with mortality among patients with COVID-19 in intensive care units in Lombardy, Italy. *Journal of the American Medical Association Internal Medicine.* 2020; **180**: 1345-1355. DOI: 10.1001/jamainternmed.2020.3539

8. **Kaplan EL, Meier P**. Nonparametric estimation from incomplete observations. *Journal of the American Statistical Association.* 1958; **53**: 457–481. DOI: 10.2307/2281868

9. **Boag JW**. Maximum likelihood estimates of the proportion of patients cured by cancer therapy. *Journal of the Royal Statistical Society, Series B (Statistical Methodology).* 1949; **11**: 15-53. http://www.jstor.org/stable/2983694

10. **Dirección Xeral de Saúde Pública** (General Directorate of Public Health), Xunta de Galicia (Autonomous Government of Galicia, NW Spain). https://www.sergas.es/Saude-publica

11. **Gude F, *et al***. Development and validation of a prognostic model based on comorbidities to predict Covid-19 severity. A population-based study. *International Journal of Epidemiology.* 2020; dyaa209. DOI: 10.1093/ije/dyaa209

12. **Maller RA, Zhou S**. *Survival analysis with long-term survivors.* Chichester, U.K.: Wiley, 1996. DOI: 10.1002/cbm.318

13. **Safari WC, López-de-Ullibarri, I, Jácome MA.** A product-limit estimator of the conditional survival function when cure status is partially known. *Biometrical Journal.* 2021; 1-22. DOI: 10.1002/bimj.202000173

14. **R Core Team**. *R: A language and environment for statistical computing.* R Foundation for Statistical Computing 2019, Vienna, Austria. https://www.R-project.org/.

15. **Beran R**. *Nonparametric regression with randomly censored survival data.* Technical Report. 1981. University of California, Berkeley

16. **Xu J, Peng Y**. Nonparametric cure rate estimation with covariates. *Canadian Journal of Statistics.* 2014; **42**: 1-17. DOI: 10.1002/cjs.11197

17. **López-Cheda A, *et al*.** Nonparametric incidence estimation and bootstrap bandwidth selection in mixture cure models. *Computational Statistics & Data Analysis.* 2017; **105**: 144-165. DOI: 10.1016/j.csda.2016.08.002

18. **López-Cheda A, Jácome MA, Cao R**. Nonparametric latency estimation for mixture cure models. *TEST.* 2017; **26**: 353-376. DOI: 10.1007/s11749-016-0515-1

19. **López-de-Ullibarri I, López-Cheda A, Jácome MA**. npcure: Nonparametric estimation in mixture cure models. 2019 R package version 0.1-4. https://CRAN.R-project.org/package=npcure

20. **Lapidus N, *et al***. Biased and unbiased estimation of the average lengths of stay in intensive care units in the COVID-19 pandemic. *Annals of Intensive Care Unit.* 2020; **10:** 135. DOI: 10.1186/s13613-020-00749-6

21. **Rees EM, *et al***. COVID-19 length of hospital stay: a systematic review and data synthesis. *BioMed Central Medicine.* 2020; **18**: 270. DOI: 10.1186/s12916-020-01726-3

22. **Wang L, *et al***. Coronavirus disease 2019 in elderly patients: Characteristics and prognostic factors based on 4-week follow-up. *Journal of Infection.* 2020; **80**: 639-645. DOI: 10.1016/j.jinf.2020.03.019

23. **Moghadas SM, *et al***. Projecting hospital utilization during the COVID-19 outbreaks in the United States. *Proceedings of the National Academy of Sciences of the United States of America.* 2020; **117**: 9122-9126. DOI: 10.1073/pnas.2004064117

24. **Li R, *et al***. Estimated demand for US hospital inpatient and intensive care unit beds for patients with COVID-19 based on comparisons with Wuhan and Guangzhou, China. *Journal of the American Medical Association Network Open.* 2020; **3**: e208297. DOI: 10.1001/jamanetworkopen.2020.8297

25. **Wood RM, *et al***. COVID-19 scenario modelling for the mitigation of capacity-dependent deaths in intensive care. *Health Care Management Science.* 2020; **23**: 315-324. DOI: 10.1007/s10729-020-09511-7

26. **Andersen PK, Keiding N**. Multi-state models for event history analysis. *Statistical Methods in Medical Research.* 2002; **11**: 91-115. DOI: 10.1191/0962280202SM276ra

27. **Meira-Machado L, de Uña-Álvarez J, Cadarso-Suárez C**. Multi-state models for the analysis of time-to-event data. *Statistical Methods in Medical Research.* 2009; **18**: 195-222. DOI: 10.1177/0962280208092301

28. **Meira-Machado L, de Uña-Álvarez J, Cadarso-Suárez C**. Nonparametric estimation of transition probabilities in a non-Markov illness-death model. *Lifetime Data Analysis.* 2006; **12**: 325-344. DOI: 10.1007/s10985-006-9009-x

29. **Meira-Machado L, de Uña-Álvarez J, Datta S**. Nonparametric estimation of conditional transition probabilities in a non-Markov illness-death model. *Computational Statistics.* 2015; **30**: 377-397. DOI: 10.1007/s00180-014-0538-6

30. **Colon ASC, Taylor JMG, Sargent DJ.** Multi-state models for colon cancer recurrence and death with a cured fraction. *Statistics in Medicine.* 2013; 33, 1750–1766. DOI: 10.1002/sim.6056

31. **Beesley LJ, Taylor JMG**. EM algorithms for fitting multistate cure models. *Biostatistics.* 2019; **20**: 416-432. DOI: 10.1093/biostatistics/kxy011

32. **Bellman RE.** *Adaptive Control Processes.* Princeton University Press, Princeton, NJ; 1961.

33. **Farewell VT**. The use of mixture models for the analysis of survival data with long-term survivors. *Biometrics.* 1982; **38**: 1041-1046.

34. **Li Q, *et al***. Early transmission dynamics in Wuhan, China, of novel coronavirus-infected pneumonia. *The New England Journal of Medicine.* 2020; **382**: 1199–1207. DOI: 10.1056/NEJMoa2001316

## Figures and Tables

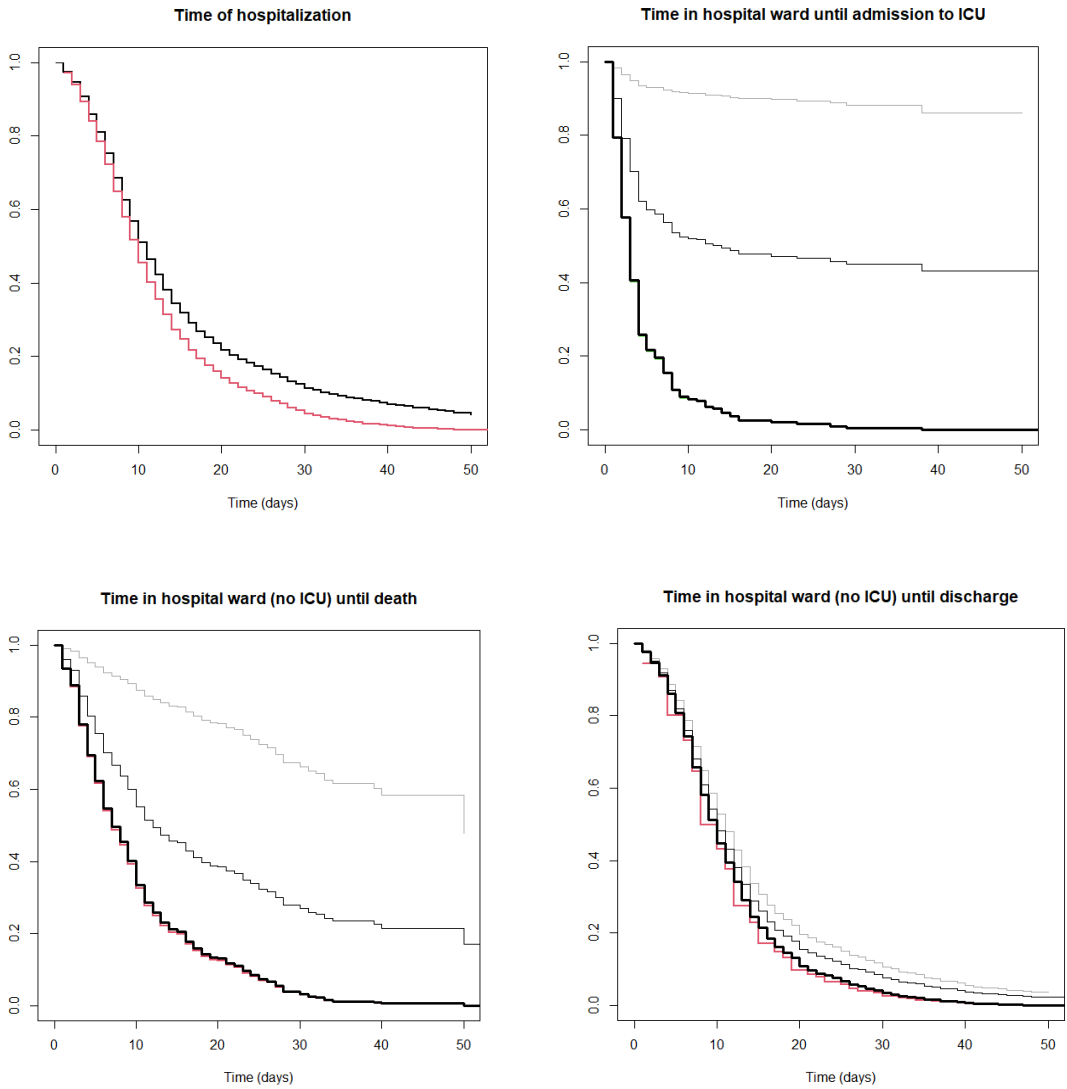

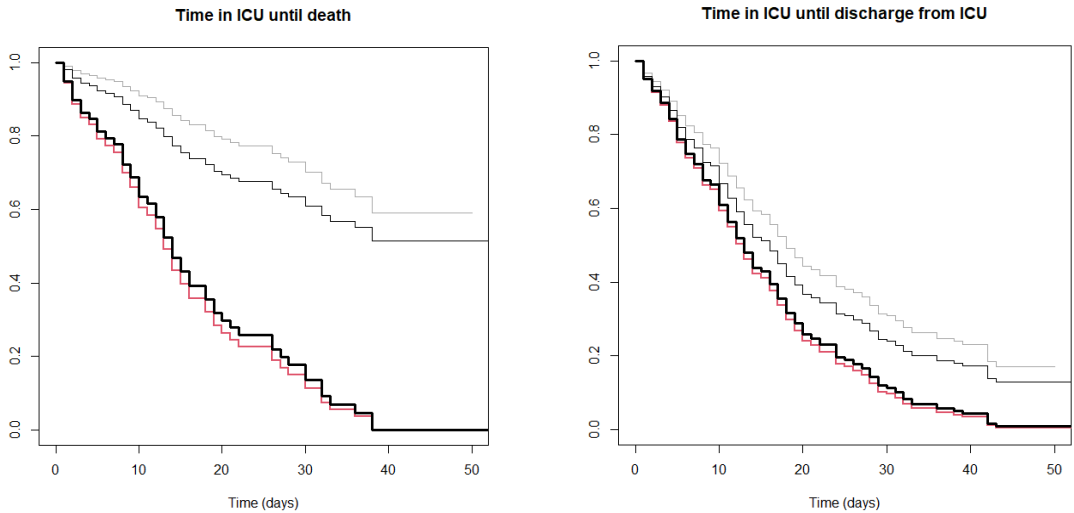

**Figure 1.** Estimates of the survival function of LoS using NP-MCM (thick black line), KM with the complete dataset (thin grey line), KM with the reduced dataset (thin black line) and the empirical E estimator (red line) for all the COVID-19 hospitalized cases (n = 2453) in Galicia (Spain), when the LoS is the time of hospitalization both in hospital ward and ICU (top left), time in hospital ward until admission to ICU (top right), time in hospital ward until death in hospital ward (middle left), time in hospital ward until discharge (middle right), time in ICU until death in ICU (bottom left) and time in ICU until discharge from ICU (bottom right). The NP-MCM and KM estimates give the same result for the time of hospitalization, and are represented with a single thick black line (top left).

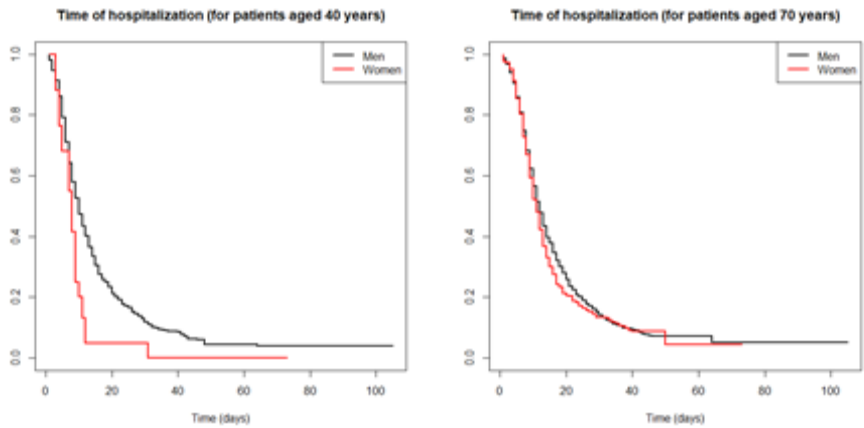

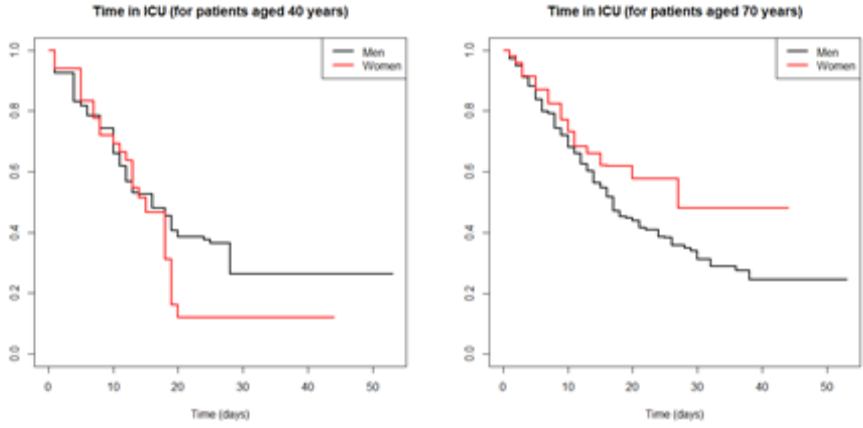

**Figure 2.** Generalized product-limit estimator [14] of the conditional survival function S(t|x) for the time of hospitalization, both in hospital ward and ICU, (top) and the time in ICU (bottom), incorporating the effect of the sex (male = black line, female = red line) and the ages 40y (left) and 70y (right) for all the COVID-19 hospitalized cases (n = 2453) in Galicia (Spain).

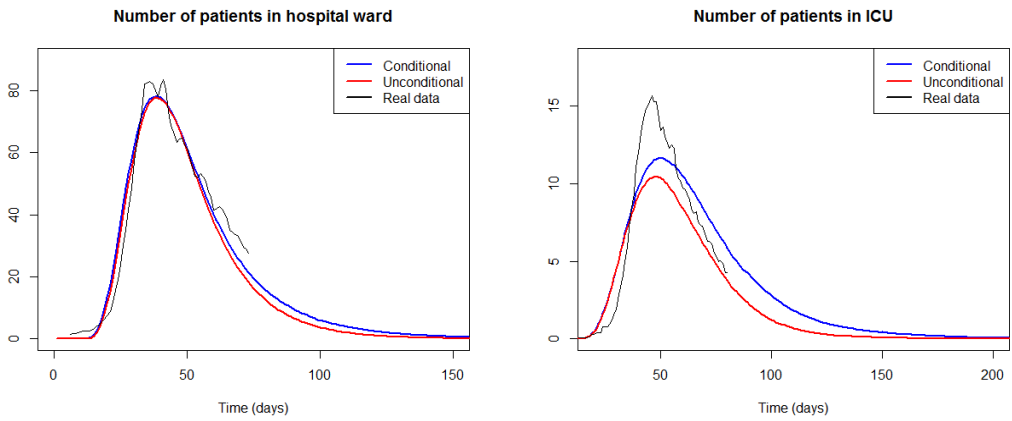

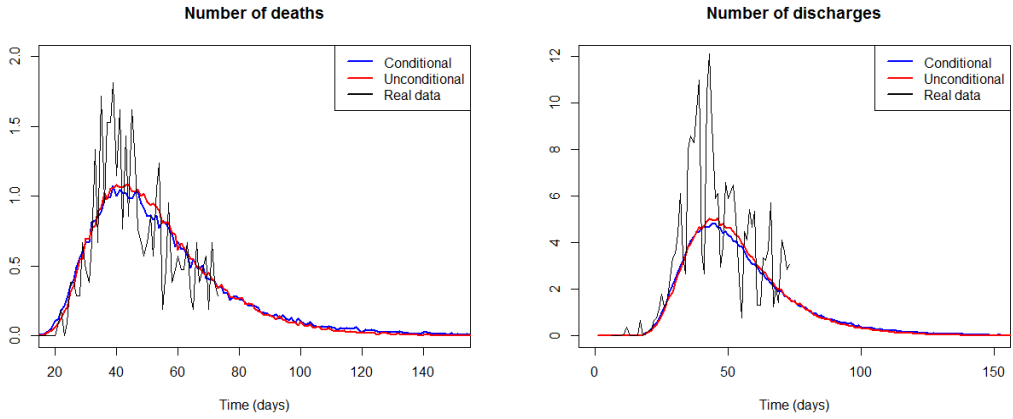

**Figure 3.** Number of patients in hospital ward (top left), ICU (top right), deaths (bottom left) and discharges (bottom right) computed from 1000 simulated COVID19 outbreaks for a period of 200 days, when the LoS are simulated depending on age and sex (blue), and unconditionally ignoring age and sex dependence (red). The real case counts of inpatients in the COVID19 dataset from March 6[th] (day 1) to May 7[th], 2020 (day 62), rescaled to N = 1000 infected people, are also included for reference (solid black line).

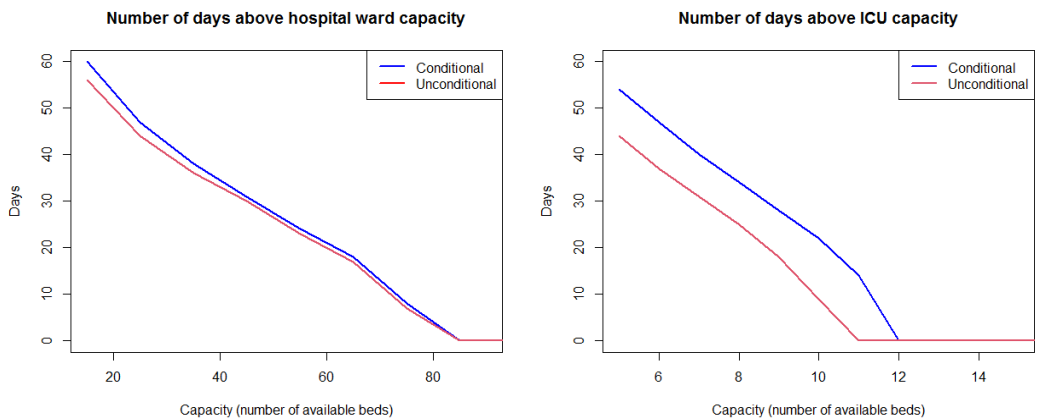

**Figure 4.** Number of days when the demand for beds is above the capacity in hospital ward (left) and ICU (right), for different possible capacities and computed from 1000 simulated COVID19 outbreaks. The demand was simulated conditionally depending on age and sex (blue), and unconditionally ignoring age and sex dependence (red).

**Table 1.** Estimated probabilities of the different medical events for the COVID-19 patients in Galicia (Spain) using NP-MCM and empirical estimators.

|  | **NP-MCM** | **Empirical** |
| --- | --- | --- |
| Need for ICU | 0.0845 | 0.0828 |
| Death in HW | 0.1561 | 0.1503 |
| Discharge from HW | 0.7953 | 0.7503 |
| Death in ICU | 0.2222 | 0.1963 |
| Discharge from ICU | 0.6820 | 0.6481 |

HW: Hospital ward; ICU: Intensive care unit